\documentclass[conference]{IEEEtran}
\IEEEoverridecommandlockouts
\usepackage{cite}
\usepackage{amsmath,amssymb,amsfonts}
\usepackage{algorithmic}
\usepackage{graphicx}
\usepackage{caption, subcaption}
\usepackage{textcomp}
\usepackage[dvipsnames]{xcolor}
\def\BibTeX{{\rm B\kern-.05em{\sc i\kern-.025em b}\kern-.08em
    T\kern-.1667em\lower.7ex\hbox{E}\kern-.125emX}}

\newcommand{\ms}{\mathcal{MS}}
\newcommand{\zz}{\mathcal{ZZ}}
\newcommand{\zx}{\mathcal{ZX}}

\usepackage{braket}
\usepackage{float}

\usepackage{qcircuit}
\usepackage{cite}
\usepackage{hyperref}
\hypersetup{
  breaklinks=true,
  colorlinks=true,
  allcolors=BlueViolet,
}
\usepackage[capitalise]{cleveref}

\Crefname{figure}{Fig.}{Figs.}
\crefname{equation}{}{}

\renewcommand{\i}{\mathrm{i}\mkern1mu} 

\newcommand{\Yb}{\textsuperscript{171}Yb\textsuperscript{+}}

\def\Superstaq/{\texttt{Superstaq}}

\begin{document}

\title{Superstaq:\\Deep Optimization of Quantum Programs
\thanks{This material is supported by the U.S. Department of Energy, Office of Science, Office of Advanced Scientific Computing Research under Award Number DE-SC0021526 and under Quantum Testbed Program Contracts No. DE-AC02-05CH11231 and DE-NA0003525. This material is based upon work supported by the U.S. Department of Energy, Office of Science, National Quantum Information Science Research Centers. Sandia National Laboratories is a multimission laboratory managed and operated by National Technology \& Engineering Solutions of Sandia, LLC, a wholly owned subsidiary of Honeywell International Inc., for the U.S. Department of Energy's National Nuclear Security Administration under contract DE-NA0003525.  This paper describes objective technical results and analysis. Any subjective views or opinions that might be expressed in the paper do not necessarily represent the views of the U.S. Department of Energy or the United States Government. We acknowledge the use of IBM Quantum services for this work. The views expressed are those of the authors, and do not reflect the official policy or position of IBM or the IBM Quantum team.
}}

\author{

\IEEEauthorblockN{
Colin Campbell\IEEEauthorrefmark{1},
Frederic T. Chong\IEEEauthorrefmark{1},
Denny Dahl\IEEEauthorrefmark{1},
Paige Frederick\IEEEauthorrefmark{1},
Palash Goiporia\IEEEauthorrefmark{1},
Pranav Gokhale\IEEEauthorrefmark{1},\\
Benjamin Hall\IEEEauthorrefmark{1},
Salahedeen Issa\IEEEauthorrefmark{1},
Eric Jones\IEEEauthorrefmark{1},
Stephanie Lee\IEEEauthorrefmark{1},
Andrew Litteken\IEEEauthorrefmark{1},
Victory Omole\IEEEauthorrefmark{1},\\
David Owusu-Antwi\IEEEauthorrefmark{1},
Michael A. Perlin\IEEEauthorrefmark{1},
Rich Rines\IEEEauthorrefmark{1},
Kaitlin N. Smith\IEEEauthorrefmark{1},
Noah Goss\IEEEauthorrefmark{2},
Akel Hashim\IEEEauthorrefmark{2},\\
Ravi Naik\IEEEauthorrefmark{2},
Ed Younis\IEEEauthorrefmark{3},
Daniel Lobser\IEEEauthorrefmark{4},
Christopher G. Yale\IEEEauthorrefmark{4}, Benchen Huang\IEEEauthorrefmark{5}, Ji Liu\IEEEauthorrefmark{6}
}

\IEEEauthorblockA{\IEEEauthorrefmark{1}Infleqtion (these authors contributed equally)}
\IEEEauthorblockA{\IEEEauthorrefmark{2}Quantum Nanoelectronics Laboratory, University of California at Berkeley}
\IEEEauthorblockA{\IEEEauthorrefmark{3}Computational Research Division, Lawrence Berkeley National Laboratory}
\IEEEauthorblockA{\IEEEauthorrefmark{4}Sandia National Laboratories}
\IEEEauthorblockA{\IEEEauthorrefmark{5}University of Chicago}
\IEEEauthorblockA{\IEEEauthorrefmark{6}Argonne National Laboratory}
}

\maketitle

\begin{abstract}
We describe \Superstaq/, a quantum software platform that optimizes the execution of quantum programs by tailoring to underlying hardware primitives. For benchmarks such as the Bernstein-Vazirani algorithm and the Qubit Coupled Cluster chemistry method, we find that deep optimization can improve program execution performance by at least 10x compared to prevailing state-of-the-art compilers. To highlight the versatility of our approach, we present results from several hardware platforms: superconducting qubits (AQT @ LBNL, IBM Quantum, Rigetti), trapped ions (QSCOUT), and neutral atoms (Infleqtion). Across all platforms, we demonstrate new levels of performance and new capabilities that are enabled by deeper integration between quantum programs and the device physics of hardware.
\end{abstract}

\begin{IEEEkeywords}
Quantum compilation, cross-layer optimization
\end{IEEEkeywords}

\section{Introduction}

Despite sustained progress in quantum hardware, there is a substantial gap to utility-scale quantum computation. On one hand, there has been consistent improvement in gate fidelities over the past decade. For instance, since the advent of superconducting transmon qubits \cite{dicarlo2009demonstration}, two-qubit gate errors have been lowered by $\sim 0.77\times$ per year \cite{zlokapa2020boundaries}. We see similar rates of progress in other qubit technologies including neutral atoms \cite{urban2009observation, graham2022multi} and trapped ions \cite{monroe1995demonstration, erhard2021entangling}. While this progress is encouraging, hardware progress alone would require at least a decade to achieve societally-useful outcomes such as simulating molecules relevant to fertilizer production \cite{lee2021even}.

\begin{figure}[t]
    \centering
    \includegraphics[width=0.42\textwidth]{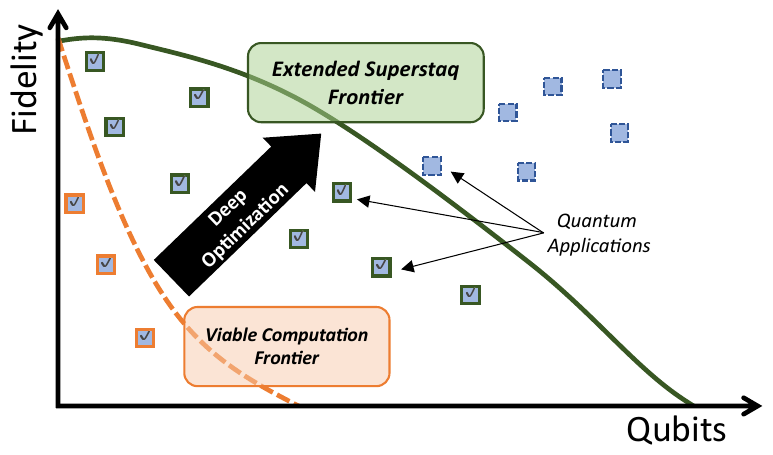}
    \caption{Deep optimization advances the frontier of which quantum programs can be run successfully.}
    \label{fig:frontier}
\end{figure}

We view quantum software as a force multiplier that can significantly shorten the timeline for utility-scale results from quantum hardware. There are compelling parallels to classical computing: the world's top computing facilities bolster their state-of-the-art hardware capabilities with significant investment in software tools such as CUDA \cite{sanders2010cuda}, Jax \cite{jax2018github}, OpenMP \cite{dagum1998openmp}, and SLURM \cite{yoo2003slurm}. Similarly, software tooling---especially for compilation---can enable users to extract better results from quantum hardware, both for near-term systems as well as for upcoming large-scale fault-tolerant computers.

In fact, we find even stronger motivation for optimized compilation in the quantum setting than in the classical setting. First, quantum resources are far more expensive than classical resources. For instance,  an error-corrected quantum NAND gate is expected to be 10 \textit{billion} times more expensive ($\sim$10 qubitseconds vs. $10^{-9}$ transistorseconds) than a classical NAND gate \cite{babbush2021focus}. Second, for foreseeable quantum computers, optimized compilation will be \textit{necessary} to bring useful applications within the boundary of achievable computations.  Lastly, applications brought within this boundary must exhibit exponential or high-degree polynomial quantum advantage \cite{babbush2021focus}, that can immediately justify high compilation costs. Thus, investment in deep compiler optimization can enable applications that are otherwise out of reach for current hardware at various scales, as depicted in Fig.~\ref{fig:frontier}.

In this spirit and with this motivation, we present \Superstaq/, a quantum software platform that invokes \textit{deep} optimization by compiling quantum programs to low levels of underlying hardware control to extract as much performance as possible. We emphasize that while compilation to standard \textit{textbook} gatesets (including work on optimal qubit mapping and routing \cite{kattemolle2023linegraph, booth2023constraint}) is addressed by existing techniques, such as open-source tools like Qiskit \cite{qiskit} and theoretical frameworks like the KAK decomposition \cite{tucci2005introduction} \& Solovay-Kitaev Theorem \cite{dawson2005solovay}, optimized compilation to lower levels remains a relatively underexplored territory. With \Superstaq/, we emphasize a full-stack approach to improving quantum compilation across multiple quantum technologies. We demonstrate that cross-layer (i.e., breaking abstractions) and device-physics-aware optimization significantly improves the performance of algorithms on quantum hardware, enabling functionality that is years ahead of the baseline hardware frontier.

\begin{table}
\caption{\label{tab:platforms} \Superstaq/ hardware targets in this paper.}
\footnotesize
\centering
\renewcommand{\arraystretch}{1.2}
\begin{tabular}{p{1.48cm}|p{2.2cm}|p{3.4cm}}
 Platform & Qubit Type & Techniques Highlighted \\ \hline \hline
 AQT @Berkeley & Supercond. (fixed-freq. transmon) & approx. synthesis, BYOG, PMW-4, qutrit decomposition \\ \hline
 IBM & Supercond. (fixed-freq. transmon)  & pulse cancellation, dynamical decoupling, fractional gates \\ \hline
 Infleqtion Hilbert & Neutral Atom \quad (Cesium) & global gate decomposition and scheduling \\ \hline
 Rigetti Aspen M-3 & Supercond. (modulated transmon) & dynamical decoupling \\ \hline
 QSCOUT @Sandia & Trapped ion (\Yb) & SWAP Mirroring, fractional gates \\ 
\end{tabular}
\end{table}

We have evaluated our optimized compilation techniques on multiple qubit technologies, in each case, compiling to the lowest level of hardware primitives accessible. In this paper, we present results from five platforms, shown in Tab.~\ref{tab:platforms}, that showcase \Superstaq/'s deep cross-layer optimization. The remainder of this paper describes the highest-leverage compiler optimizations, with experimental validation for each. Section~\ref{sec:optimized_decomposition} presents results that emerge from optimization of quantum programs to lower abstraction levels (native gateset or pulses) than typically considered. Section~\ref{sec:dynamical_decoupling} presents \Superstaq/'s integration of dynamical decoupling, which we find to be a particularly profitable technique for mitigating noise. Finally Section~\ref{sec:advanced_mapping} presents the \textit{star-to-line} routing, a simple-but-effective technique that bridges the gap from low-connectivity hardware to typical applications.

Our software can be accessed through open-source Python frontends, \texttt{cirq-superstaq} and \texttt{qiskit-superstaq}, available through PyPI via \texttt{pip install}. Both packages include comprehensive libraries of custom gates \href{https://github.com/SupertechLabs/client-superstaq/blob/main/cirq-superstaq/cirq_superstaq/ops/qubit_gates.py}{\texttt{css.ops.qubit\_gates}}, \href{https://github.com/SupertechLabs/client-superstaq/blob/main/cirq-superstaq/cirq_superstaq/ops/qudit_gates.py}{\texttt{css.ops.qudit\_gates}}, and \href{https://github.com/SupertechLabs/client-superstaq/blob/main/qiskit-superstaq/qiskit_superstaq/custom_gates.py}{\texttt{qss.custom\_gates}}, that support \Superstaq/'s interaction with the low-level hardware primitives referenced in this paper.

\section{Optimized Decomposition} \label{sec:optimized_decomposition}

\Superstaq/'s approach to optimized decomposition begins by defining the native gateset of the hardware. We define a device's ``native gates'' to be the smallest operation that has a known expected unitary. For example, on many hardware platforms, multi-qubit gates are implemented by an echoed interaction, whereby interaction Hamiltonians are applied in two half-steps: positive and negative directions. The isolated half-steps do not have a known expected unitary (due to an uncharacterized error term), but the composite sequence does. We therefore consider the full echoed sequence as the level of abstraction that forms our native gateset. We expand on this with the concrete example of Echoed Cross-Resonance gates on IBM hardware in the next subsection.

By first identifying the native gateset of each device, \Superstaq/ is able to tailor circuit decompositions to exploit the full capabilities of the hardware. It combines a number of novel, platform-aware decomposition and optimization strategies, as well as off-the-shelf optimizers available in packages like Cirq, Qiskit, and BQSKit \cite{bqskit2021}. In future iterations we also plan to incorporate \textit{superoptimization} passes \cite{xuquartz} to further ensure the optimal translation to a given gateset.

This section explores a few of the specific optimized decomposition strategies employed by \Superstaq/ for a variety of hardware platforms. In each case we find that this cross-layer approach extends the capabilities of the device beyond what would be possible with a hardware-agnostic approach.



\subsection{Superconducting Gateset on IBM (Echoed Cross-Resonance)}

We begin by describing \Superstaq/'s optimizations for IBM superconducting hardware. At a high level, our compiler performs three key steps: (a) qubit mapping, (b) decomposing into native gatesets to surface opportunities for cross-gate pulse cancellation, and then (c) merging single-qubit gates to minimize rotation angle and prefer use of virtual $R_z$ rotations over physical $R_x$ rotations.

On current IBM hardware, two-qubit interactions are performed with the cross-resonance pulse  \cite{gambetta2020crossreso}, with Hamiltonian:
\begin{equation*}
\hat{H} = \omega_{ix} \frac{\hat{I}\hat{X}}{2} + \omega_{iz} \frac{\hat{I}\hat{Z}}{2} + \omega_{zi} \frac{\hat{Z}\hat{I}}{2} + \omega_{zx} \frac{\hat{Z}\hat{X}}{2} + \omega_{zz} \frac{\hat{Z}\hat{Z}}{2}.
\end{equation*}
The cross-resonance interaction amounts to driving the control qubit at the natural frequency of the target qubit. On IBM devices, the $\omega_{zx}$ frequency is used to calibrate an entangling gate. By applying this Hamiltonian in an echoed fashion with both positive and negative drives, with an $X$ gate in between, the resulting unitary implements only the $\hat{Z}\hat{X}$ interaction, plus a side effect due to the $X$ gate to reduce error \cite{Sundaresan_2020}. This is the native two-qubit gate we target with \Superstaq/.

We demonstrate \Superstaq/ optimizing a Qubit-Coupled Cluster circuit~\cite{huang2023qcc_ansatz, ryabinkin2018qubit}, useful for quantum chemistry applications. The core subcircuit of QCC consists of an $R_z$ operation sandwiched by four total CX gates, two pairs each staggered in a “ladder” formation on either side \cite{gui2020term}, as seen in \cref{subfig:uccsd}. \Superstaq/ compiles this circuit using (a) cross-gate pulse cancellations \cite{gokhale2021faster} and (b) as-late-as-possible scheduling. Firstly, \Superstaq/ leverages cross-gate pulse cancellations by decomposing the CX gate into single-qubit gates and an echoed cross-resonance. Decomposing the CX into its native gates reveals further single-qubit gate decompositions that are unavailable to Qiskit, even at its highest optimization level.

\begin{figure}[h]
    \centering
    \begin{subfigure}{\linewidth}
        \centering
        \includegraphics[scale=0.35]{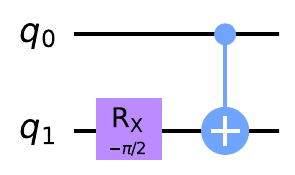}
        \includegraphics[scale=0.35]{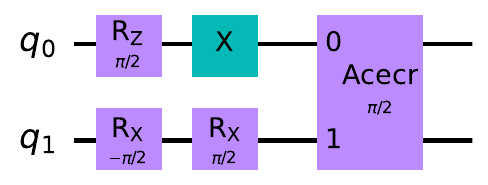}
        \includegraphics[scale=0.35]{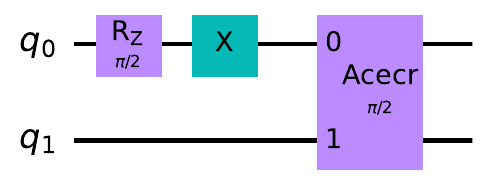}
        \caption{\Superstaq/ optimization.}
        \label{subfig:x-cx-qc}
    \end{subfigure} \\[.5em]
    \begin{subfigure}{\linewidth}
        \centering
        \includegraphics[scale=0.15]{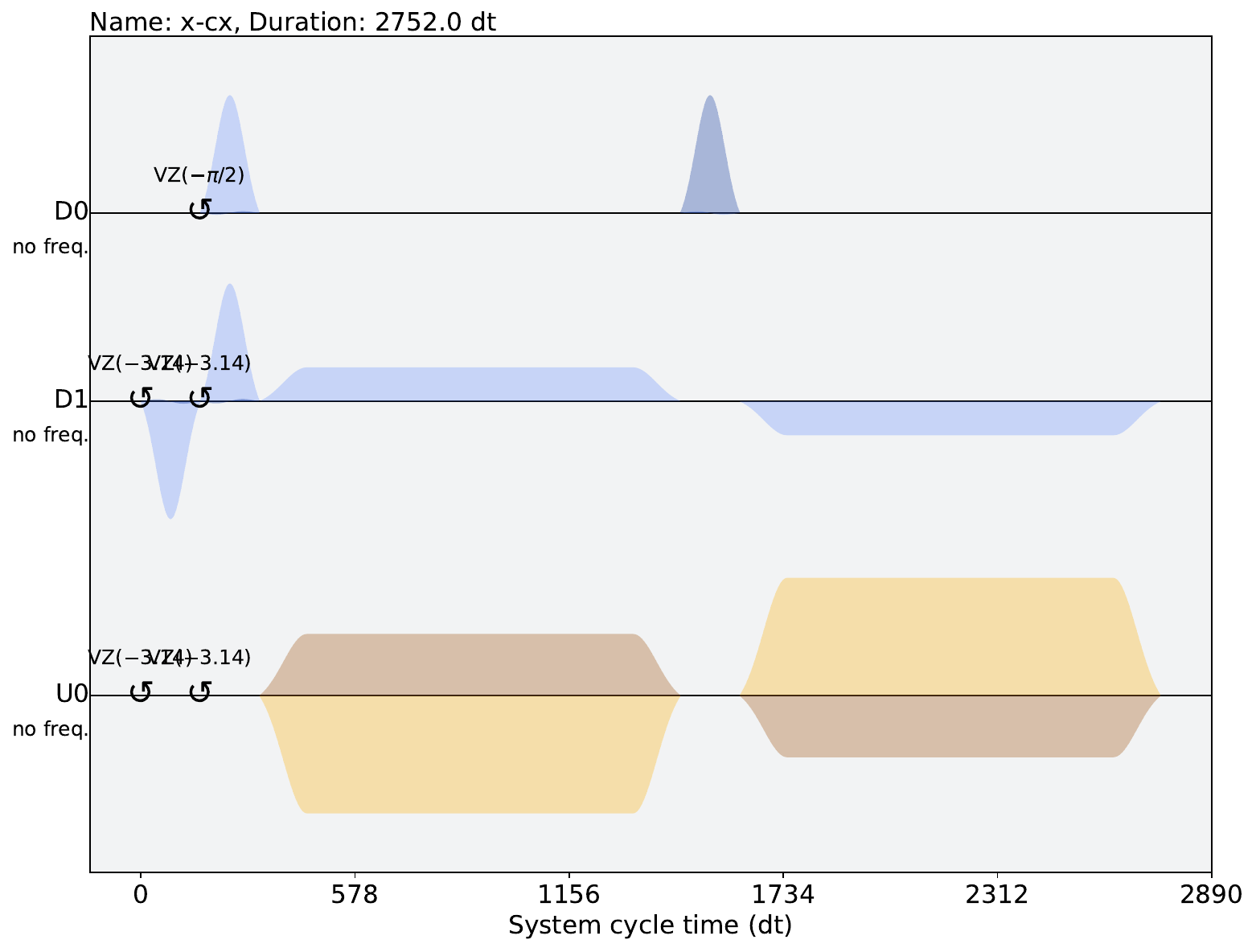}
        \includegraphics[scale=0.15]{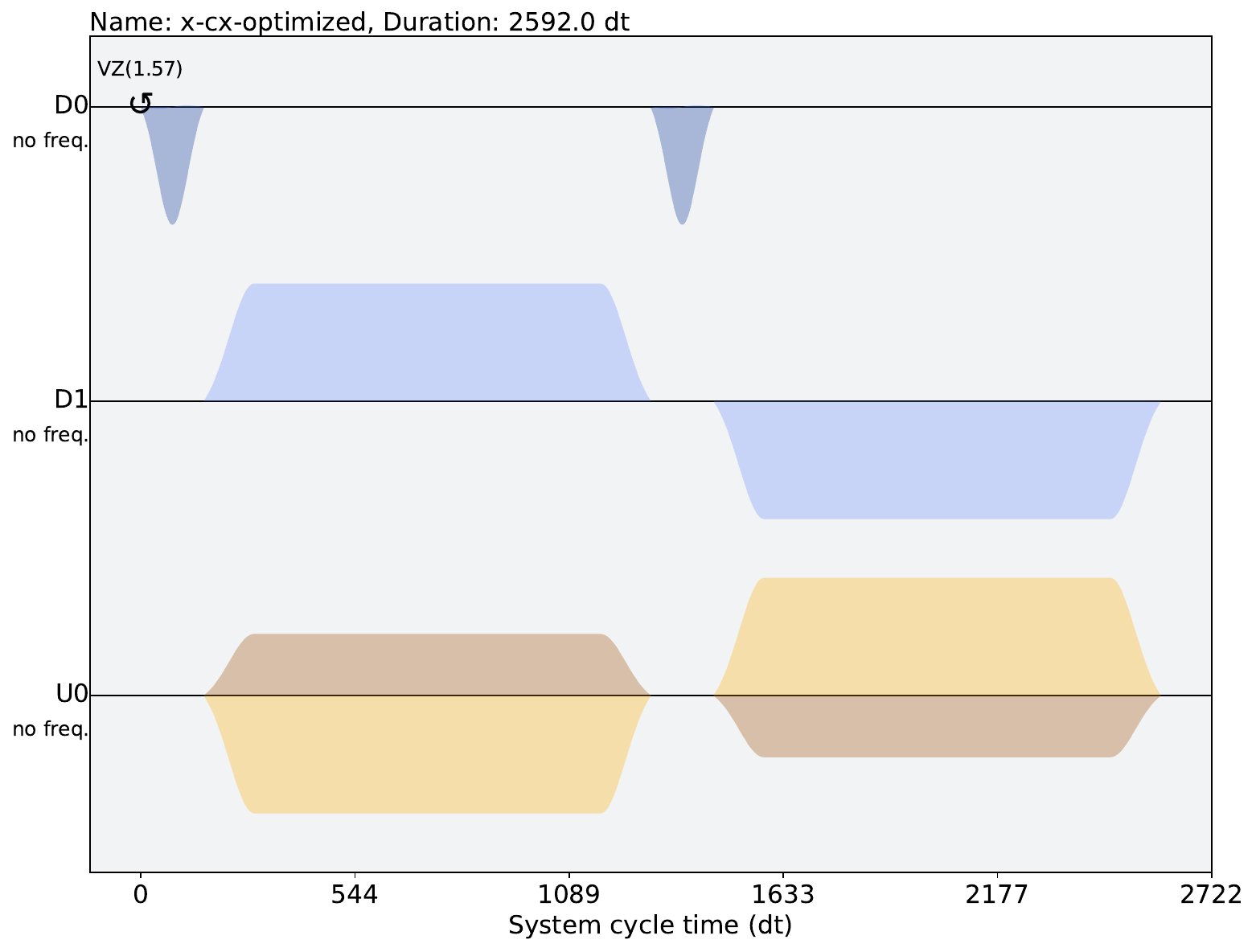}
        \caption{Comparison of pulse schedules generated with Qiskit (\texttt{optimization\_level=3}) and \Superstaq/.}
        \label{subfig:x-cx-sched}
    \end{subfigure} \\[.5em]
    \caption{(a) A circuit consisting of a $X$-rotation followed by a CX gate (left), along with the subsequent decomposition into native gates (center) and cross-gate cancellation carried out by \Superstaq/ (right). (b) The Qiskit schedule of the original circuit 
    (left, $2752\text{dt} \approx 612\text{ns}$) and the \Superstaq/ optimized schedule (right, $2592\text{dt} \approx 576\text{ns}$),
    achieving a savings of $160\text{dt} \approx 36\text{ns}$, exactly the duration of a single-qubit gate. (Note: the amplitude scale for each channel is normalized.)}
    \label{fig:x-cx}
\end{figure}

\begin{figure}
    \centering
    \begin{subfigure}{\linewidth}
        \centering
        \includegraphics[scale=0.3]{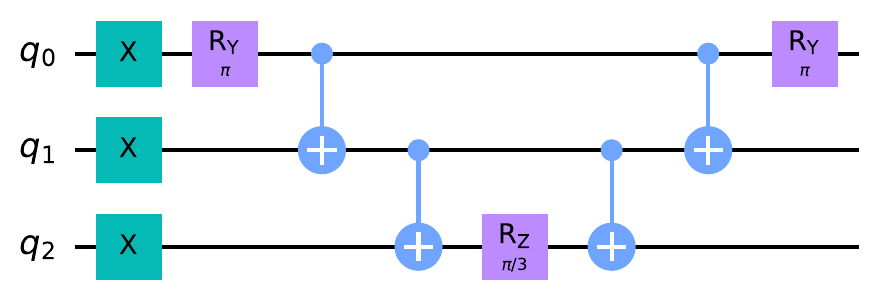}
        \caption{Exemplar core subcircuit for QCC}
        \label{subfig:uccsd}
    \end{subfigure} \\[.5em]
    \begin{subfigure}{\linewidth}
        \centering
        \includegraphics[scale=0.14]{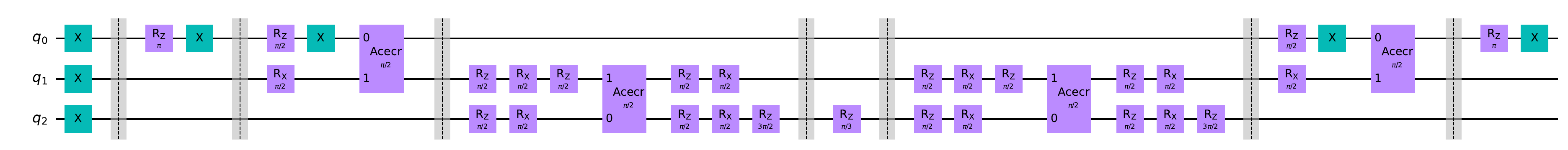}
        \caption{Decomposed core QCC subcircuit}
        \label{subfig:uccsd-decomp}
    \end{subfigure} \\[.5em]
    \begin{subfigure}{\linewidth}
        \centering
        \includegraphics[scale=0.22]{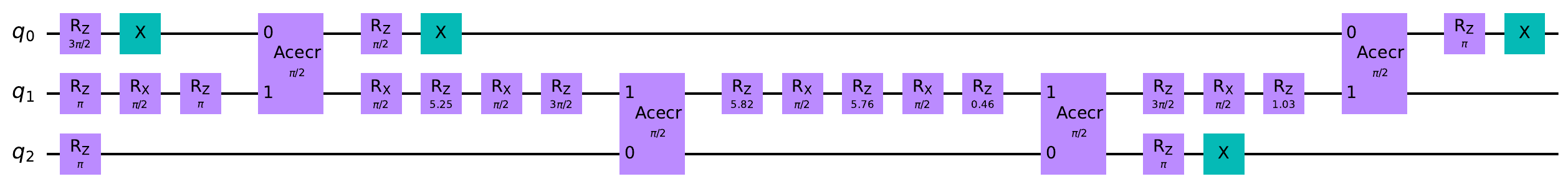}
        \caption{\Superstaq/ optimized core QCC subcircuit}
        \label{subfig:uccsd-opt}
    \end{subfigure} \\[.5em]
    \caption{\Superstaq/ optimization of the Qubit Coupled Cluster ansatz (QCC).}
    \label{fig:uccsd-optimization}
\end{figure}

Before describing the full optimization of the QCC circuit, first we’ll describe a representative warm-up example involving an $R_x$ gate followed by a CX in \cref{fig:x-cx}. Qiskit will execute this circuit directly. \Superstaq/ decomposes the CX into native gates: (1) an $R_z(\pi/2)$ gate followed by an $X$ gate on the first qubit, (2) an $R_x(\pi/2)$ on the second qubit, and (3) an echoed cross-resonance gate interacting the first and second qubits (\cref{subfig:x-cx-qc}) \cite{bowman2022hardware}. \Superstaq/ then cancels the $R_x(\pi/2)$ gate with the preceding $R_x(-\pi/2)$ gate on the second qubit, reducing the duration of the resulting pulse schedule executed on hardware (\cref{subfig:x-cx-sched}).

\begin{figure}
    \centering
    \includegraphics[width=0.49\textwidth]{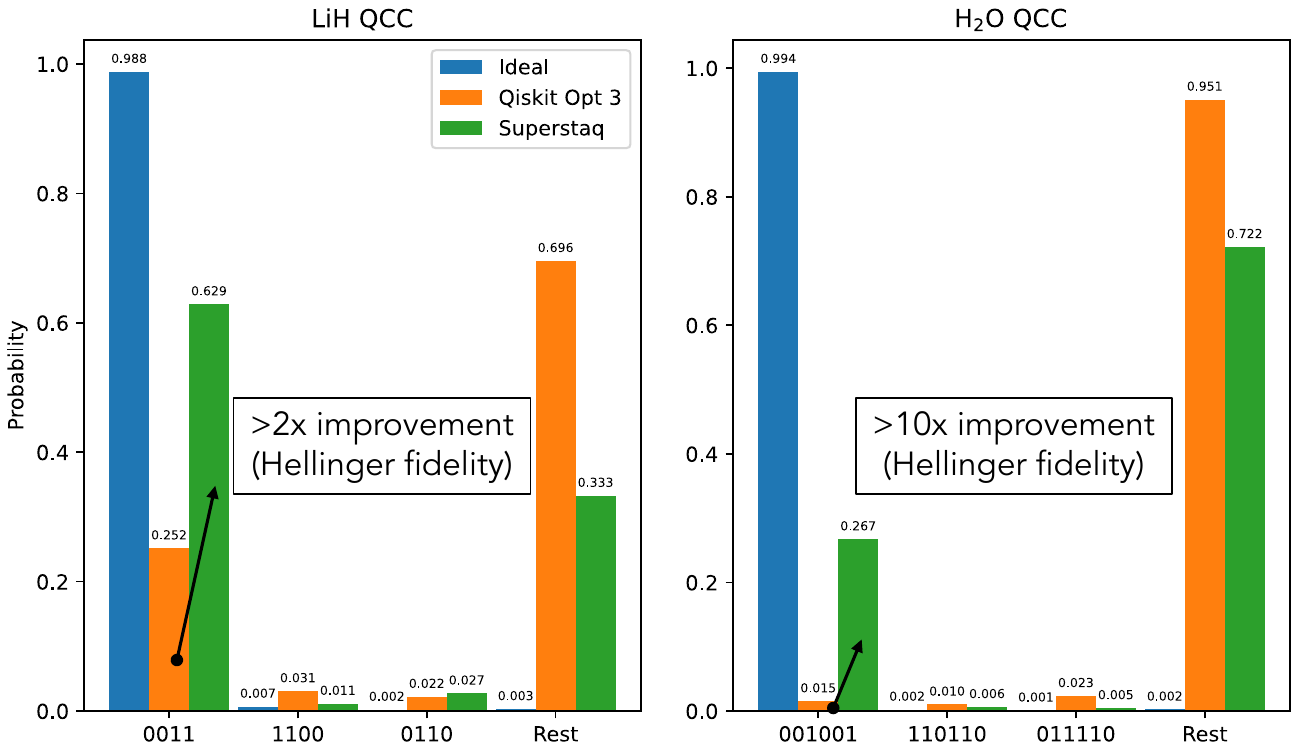}
    \caption{Qubit Coupled Cluster experimental results on ibm\_lagos device for LiH (4 qubits, 31 two-qubit gates) and H\textsubscript{2}O (6 qubits, 57 two-qubit gates), 32k shots per bar. \texttt{Superstaq} improves Hellinger fidelity 28.6\% $\rightarrow$ 66.3\% (2.3x) and 2.1\% $\rightarrow$ 28.2\% (13x) respectively. Ideal results are almost (but not completely) single-peaked.}
    \label{fig:qcc_ss_results}
\end{figure}

Now we again consider the core QCC subcircuit. We first decompose the CXs in the subcircuit to obtain the result shown in \cref{subfig:uccsd-decomp}. In this initial decomposition, \Superstaq/ reveals a number of single-qubit gate cancellation opportunities similar to as seen in the warm-up example. In particular we see instances of the following gate sequences:
\begin{itemize}
\small
    \item \texttt{X-Rz($\pi$)-X-Rz($\pi/2$)-X} on $q_0$
    \item \texttt{X-Rx($\pi/2$)} on $q_1$
    \item \texttt{Rz($\pi/2$)-Rx($\pi/2$)-Rz($\pi/2$)-Rx($\pi/2$)-Rz($\pi/2$)} on $q_1$
    \item \texttt{Rz($\pi/2$)-Rx($\pi/2$)-Rx($\pi/2$)} on $q_1$
    \item \texttt{X-Rz($\pi/2$)-Rx($\pi/2$)} on $q_2$
    \item \texttt{Rz($\pi/2$)-Rx($\pi$/2)-Rz(3$\pi/2$)-Rz($\pi/3$)-Rz($\pi/2$)} \texttt{-Rx($\pi/2$)} on $q_2$
\end{itemize}
Each of these sequences can be recombined into either (a) shorter sequences consisting of $R_z$ and $R_x$ gates, or (b) sequences that reduce the use of $R_x$ gates in favor of $R_z$ gates. In addition to the reduction in number of gates with (a), because $R_z$ gates on IBM devices are virtual \cite{McKay_2017}, (b) achieves additional savings. The resultant re-compiled sequences are, respectively:
\begin{itemize}
\small
    \item \texttt{Rz($3\pi/2$)-X} on $q_0$
    \item \texttt{Rz($\pi$)-Rx($\pi/2$)-Rz($\pi$)} on $q_1$
    \item \texttt{Rz($\pi$)-Rx($\pi/2$)-Rz($\pi$)} on $q_1$
    \item \texttt{Rz($\pi/2$)-X} on $q_1$
    \item \texttt{Rz($\pi/2$)-Rx($\pi/2$)-Rz($\pi/2$)} on $q_2$
    \item \texttt{Rz($3\pi/2$)-Rx($\pi/2$)-Rz($5\pi/3$)-Rx($\pi/2$)-Rz($\pi$)} on $q_2$
\end{itemize}
The final optimized result is shown in \cref{subfig:uccsd-opt}. We achieve a savings of $320\text{dt} \approx 71\text{ns}$ in pulse duration for the QCC subcircuit. As shown in Fig.~\ref{fig:qcc_ss_results}, we executed the \texttt{Superstaq}-optimized QCC circuits on IBM quantum hardware. \texttt{Superstaq} increases Hellinger fidelity by 2.3x and 13x compared to Qiskit's highest optimization, level-3.

This is a promising start, but we can go even further. Note that the cross-resonance implements a $\zx(\pi/2)$, but with pulse-stretching, \Superstaq/ provides an $\operatorname{AceCR}(\theta)$ gate implementing a parametrized $\zx(\theta)$. The parametric cross-resonance gives users the ability to compile to $\zx(\theta)$ directly in hardware, rather than decompose it into a native gate sequence. For example, the parametric cross-resonance can replace the standard cross-resonance in a CX implementation, turning it into a fractional gate $(\mathrm{CX})^\alpha$ for $\alpha = \theta / 2\pi$ that results in shorter pulse sequences that are more accurate to evaluate, thus allowing faster convergence for variational algorithms \cite{meitei2021gatefree, earnest2021pulse, gokhale2020optimized}.

\begin{figure}[h]
    \centering
    \begin{subfigure}{\linewidth}
        \centering
        \includegraphics[scale=0.4]{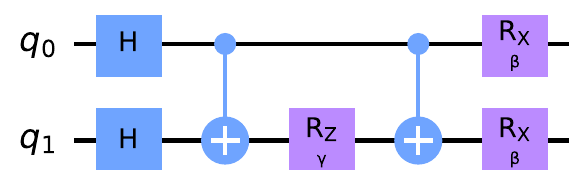}
        \includegraphics[scale=0.4]{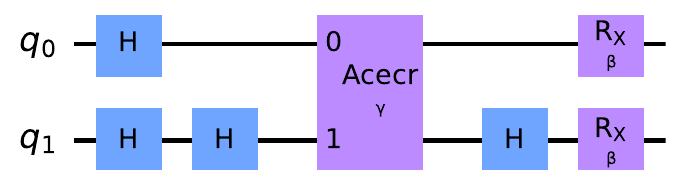}
        \caption{Two-qubit QAOA ansatz.}
        \label{subfig:qaoa-toy}
    \end{subfigure} \\[.5em]
    \begin{subfigure}{\linewidth}
        \includegraphics[scale=0.6]{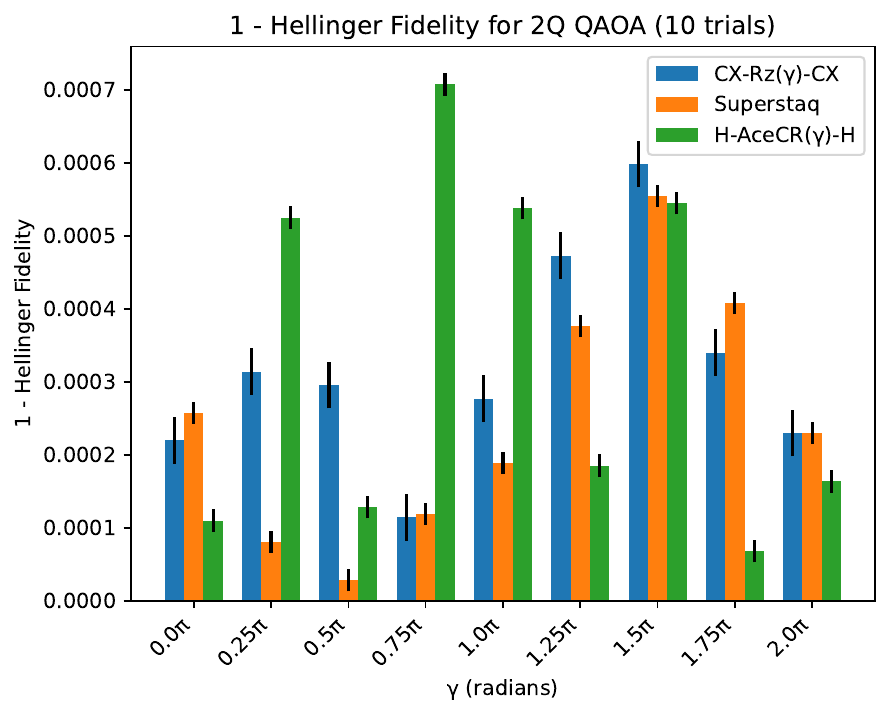}
        \caption{1 - Hellinger fidelity results for $\gamma\in[0, 2\pi]$ and $\beta=\pi/2$ fixed.}
        \label{subfig:hellinger-qaoa}
    \end{subfigure} \\[.5em]
    \caption{(a) Qiskit ZZ decomposition (two CX gates) vs. direct (i.e., with AceCR) ZZ decomposition. (b) Plots comparing error between Qiskit ZZ decomposition, standard \Superstaq/, and \Superstaq/ AceCR for $\gamma\in[0, 2\pi]$.}
    \label{fig:qaoa}
\end{figure}

We now consider the Quantum Approximate Optimization Algorithm (QAOA), a variational quantum algorithm used to solve combinatorial optimization \cite{farhi2014quantum}. The QAOA ansatz consists of three stages: (1) a layer of Hadamards putting each qubit into an equal superposition, (2) a ZZ-type interaction between desired pairs of qubits based on the problem Hamiltonian, and (3) a final mixing layer of $R_x$ rotation gates. In general, this ansatz can be extended to larger depth for more accurate convergence by repeating the final two stages in sequence $p$ times. We consider a two-qubit toy model of the $p=1$ QAOA ansatz, as shown in the left hand side of \cref{subfig:qaoa-toy}, consisting of the standard decomposition for a single ZZ-type interaction. Out of the box, \Superstaq/ optimizes this decomposition via cross-gate pulse cancellations between each CX gate and the surrounding single-qubit gates, similar to in \cref{fig:x-cx,fig:uccsd-optimization}. Additionally, \Superstaq/'s aforementioned isolation of the cross-resonance at the gate-level allows implementing entangling gates other than the ZX-type interaction via phase kickback, of the form $\{\sigma_i\sigma_j\mid i,j\in x, y, z\}$ for Pauli operators $\sigma_i$ (see Figure 4 in \cite{danin2023procedure}). For example, the ZZ interaction in \cref{subfig:qaoa-toy} can be implemented directly with the $\mathrm{AceCR}(\theta)$, as shown in the right hand side of \cref{subfig:qaoa-toy}, requiring fewer two-qubit native gates. We compare the gate errors (1 - Hellinger fidelities) for Qiskit, \Superstaq/'s standard optimization, and the direct $\operatorname{AceCR}(\theta)$ implementation facilitated by Superstaq in \cref{subfig:hellinger-qaoa} and we see that except for $\gamma=7\pi/4$, \Superstaq/ either ties or beats Qiskit i.e. the orange bar is tied with or lower than the blue bar. The error bar is the standard error of the mean $\frac{\sigma}{\sqrt{N}}$ for $N$ trials, and a tie occurs when the error bars intersect. We note that while Qiskit beats the direct $\operatorname{AceCR}(\theta)$ implementation for $\gamma \in \{0.25\pi, 0.75\pi, \pi\}$, future iterations of Superstaq will address this with custom pulse-shaping of the $\operatorname{CR}$, rather than relying on Qiskit's calibration.

\subsection{Neutral Atom Gateset}

Here we discuss \Superstaq/'s compilation to Infleqtion's Hilbert QPU, which has a native gateset comprising:
\begin{itemize}
  \item Two-qubit CZ gates.
  \item Single-qubit rotations of the form $R_z(\theta)=e^{-\i\theta Z_j/2}$, where $Z_j$ is the Pauli-$Z$ operator for qubit $j$.
  \item Global rotations of the form $GR_\phi(\theta)=e^{-\i\theta S_\phi}$, where the spin operator $S_\phi = \frac12\sum_j(\cos(\phi) X_j + \sin(\phi) Y_j)$ generates homogeneous rotations of all qubits in the $X$-$Y$ plane, and $X_j, Y_j$ are Pauli operators for qubit $j$.
\end{itemize}
The error budget of a quantum computation is typically dominated by two-qubit gates (e.g.,~CZ gates), and the neutral atom architecture is no exception.
Leveraging a large volume of literature on minimizing two-qubit gate costs, the \Superstaq/ compiler first uses decompositions available in \texttt{Cirq} \cite{cirq_developers_2022_7465577} to convert an arbitrary circuit into a gateset of CZ gates and arbitrary single-qubit gates.
However, Hilbert's gateset has the nonstandard feature that a global $GR$ gate addressing all qubits is in turn required to implement an arbitrary single-qubit gate.
\Superstaq/'s Hilbert compiler therefore optimizes to the capabilities and limitations of the $GR$ gate.

While CZ and $R_z$ gates on the neutral atom architecture occur on  time scales measured in hundreds of nanoseconds, the $GR_\phi(\theta)$ gate takes a time that is directly proportional to its \textit{pulse area} $\theta$, and takes a few microseconds when $\theta=\pi$.
Minimizing the usage of $GR$ gates therefore directly translates into decreased circuit runtimes in this neutral atom architecture.

When decomposing arbitrary single-qubit gates into $GR$ and $R_z$ gates, it is possible to ``recycle'' $GR$ gates and decompose single-qubit gates in parallel.
The \Superstaq/ compiler therefore has two basic strategies for minimizing $GR$ gate usage:
(a) scheduling single-qubit gates in such a way as to minimize the number of times that a collection of single-qubit gates must be decomposed into $GR$ and $R_z$ gates, and
(b) decomposing any given collection of single-qubit gates in such a way as to minimize the required pulse area of $GR$ gates in the decomposition.

\begin{figure}
    \centering
    \begin{subfigure}{\linewidth}
      \centering
      \includegraphics[scale=0.9]{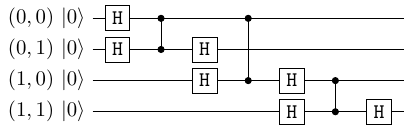}
      \caption{A GHZ circuit for four qubits at the corners of a square.}
    \end{subfigure} \\[.5em]
    \begin{subfigure}{\linewidth}
      \centering
      \includegraphics[scale=0.9]{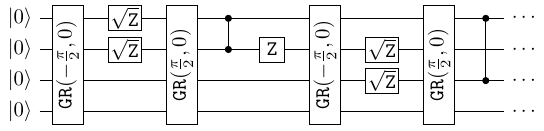}
      \includegraphics[scale=0.9]{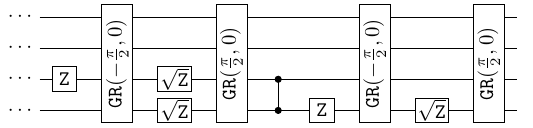}
      \caption{A straightforward compilation of the GHZ circuit for Hilbert.}
    \end{subfigure} \\[.5em]
    \begin{subfigure}{\linewidth}
      \centering
      \includegraphics[scale=0.9]{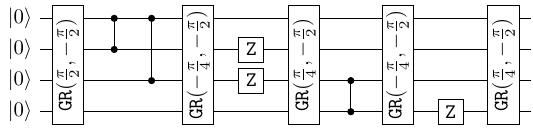}
      \caption{The same circuit compiled using \Superstaq/.}
    \end{subfigure}
    \caption{(a) A GHZ circuit for four qubits at the corners of a square, decomposed down to CZ gates and single-qubit Hadamards.
    (b) A straightforward decomposition of the GHZ circuit, in which each Hadamard is decomposed as ${H} = e^{-i\pi/4} Z \cdot {GR}_x(\frac{\pi}{2}) \cdot \sqrt{Z} \cdot {GR}_x(-\frac{\pi}{2})$, and pairs of Hadamards are decomposed using the same $GR$ gates.
    (c) The same circuit decomposed using \Superstaq/, which optimizes single-gate parallelization and minimizes $GR$ gate usage.}
    \label{fig:GHZ_circuits}
\end{figure}

Single-qubit gates are scheduled by a greedy algorithm that iterates over the operations in a circuit in topological order to collect maximal sets of single-qubit gates that can be executed in parallel \cite{nottingham2023decomposing}.
This algorithm is both efficient and optimal in reducing the number of single-qubit gate collections that must be decomposed into global gates.

Ignoring global phases, each single-qubit gate can be written in the form $Z^{z+a} X^x Z^{-a}$ with $x\in[0,1]$, where if $x=0$ the gate can simply be merged into the next single-qubit gate with $x\ne0$ on the same qubit.
The minimum global pulse area required to implement such a gate is $x \pi$, and the minimum global pulse area required to implement a collection $\mathcal{C}$ of such gates in parallel is $\max_{j\in\mathcal{C}} x_j \pi$.
\Superstaq/ achieves this minimum with a decomposition of the gates in $\mathcal{C}$ into local $R_z$ gates and two $GR$ gates that are mutual inverses, and each have a pulse area of $\min_{j\in\mathcal{C}} x_j \pi/2$.
The phase $\phi$ of the $GR$ gates in this decomposition is chosen in such a way as to minimize the number of $R_z$ gates in the compiled circuit  \cite{nottingham2023decomposing}.

As a final point, we note that the assignment of single-qubit gates to a minimal collection of parallelizable gates is an over-determined problem.
After an initial scheduling pass and prior to decomposing single-qubit gates, \Superstaq/ therefore additionally assigns single-qubit gates to collections in such a away as to minimize the net $GR$ pulse area of their decompositions.

\Cref{fig:GHZ_circuits} shows an example of equivalent circuits for preparing a four-qubit GHZ state, compiled either using off-the-shelf methods in \texttt{Cirq} (see figure caption), or using \Superstaq/.
In this example, the techniques in \Superstaq/ reduce the $R_z$ gate count from 10 to 3, the $GR$ gate count from 8 to 5, and the net $GR$ pulse area from $4\pi$ to $1.5\pi$.
Using an experimentally motivated \cite{graham2022demonstration} depolarizing noise model with a single-qubit SPAM fidelity of 0.98, $GR$ fidelity of 0.999 (per qubit), $R_z$ fidelity of 0.99, and a CZ fidelity of 0.96, we estimate that the circuits in \cref{fig:GHZ_circuits} should prepare a four-qubit GHZ state with respective fidelities of 0.71 (straightforward compilation) and 0.78 (\Superstaq/ compilation).
Remarkably, this back-of-the-envelope estimate is in quantitative agreement with our experimental trials on Hilbert, which use population measurements and parity oscillations to extract GHZ fidelities of 0.726(9) and 0.782(8).
This agreement should not to be taken too seriously due to the simplicity of the noise model -- $R_z$ and CZ gates are known to be dominated by phase errors, for example.
Nonetheless, as a rough comparison we find that the same improvement to the GHZ fidelity of the straightforwardly compiled circuit can be achieved by improving the CZ fidelity from 0.96 to 0.99.
While this comparison should not be taken literally, it illustrates the benefit of an improved compiler, which complements hardware improvements.
Note that quantum errors compound multiplicatively throughout a circuit, such that the benefits of an improved compiler grow with increasing circuit size.


\subsection{Trapped Ion Gateset}

%
%
We next describe \Superstaq/ optimization for QSCOUT \cite{clark2021qscout}, the trapped ion testbed at Sandia Lab. QSCOUT's gateset is generated via Raman transitions of a hyperfine `clock' transition of {\Yb} ions and consists of continuously parameterized single-qubit rotations, ${R}_{\phi}(\theta)$ about any axis in the $X$/$Y$ plane, a virtual $Z(\theta)$ gate, and a continuously parameterized M{\o}lmer-S{\o}rensen ($\ms_{\phi}(\theta)$) gate.

\subsubsection{Entangling Basis Compilation}

The $\ms_{\phi}(\theta)$ interaction is an $XX$-type interaction of the form ${XX}(\theta) = e^{-i\frac{\theta}{2} \sigma_{X} \otimes \sigma_{X}}$, and we adjust $\theta$ via the amplitude of one of the Raman beams \cite{shaffer2023rav}.
However, the bare interaction, ${MS}^{cu}_{\phi}(\theta)$, requires Raman transitions generated by counter-propagating beams for motional sensitivity, while the single-qubit gates use a single beam with two Raman tones, a co-propagating configuration. To mitigate phase instabilities that occur when mixing gates of differing propagation, a bare ${MS}_{xx}^{cu}(\theta)$ is sandwiched first by counter-propagating rotation gates, ${R}_y^{cu}(\pm\pi/2)$, converting the interaction into a $ZZ$-type interaction\cite{lee2005phasegates}, $\zz(\theta)$, and then a set of co-propagating rotation gates ${R}_{\phi+\pi/2}^{co}(\pm\pi/2)$, converting it back into an $XX$-type interaction, $\ms_{\phi}(\theta)$, in which the phase $\phi$ is encoded in the co-propagating rotation gates. \Superstaq/ uses just the internal $\zz(\theta)$ portion of the {MS} gate when decomposing arbitrary two-qubit operations, requiring fewer single-qubit operations than would the standard compilation. 

This is shown below, where,
\begin{equation*}
\Qcircuit @C=1em @R=.7em {
& \multigate{1}{\mathcal{MS}_\phi(\theta)} & \qw  \\
& \ghost{\mathcal{MS}_\phi(\theta)} & \qw }
\end{equation*}
is equal to,
\begin{equation*}
\Qcircuit @C=.35em @R=.35em {
& \gate{{R}_{\phi+\frac{\pi}{2}}^{co}(\frac{\pi}{2})} & \gate{{R}_y^{cu}(\frac{\pi}{2})} & \multigate{1}{{MS}_{xx}^{cu}(\theta)} &  \gate{{R}_y^{cu}(-\frac{\pi}{2})} & \gate{{R}_{\phi+\frac{\pi}{2}}^{co}(-\frac{\pi}{2})} & \qw  \\
& \gate{{R}_{\phi+\frac{\pi}{2}}^{co}(\frac{\pi}{2})} & \gate{{R}_y^{cu}(\frac{\pi}{2})} & \ghost{{MS}_{xx}^{cu}(\theta)} &  \gate{{R}_y^{cu}(-\frac{\pi}{2})} & \gate{{R}_{\phi+\frac{\pi}{2}}^{co}(-\frac{\pi}{2})} & \qw \gategroup{1}{3}{2}{5}{.4em}{--} \\
\vspace{3cm} \\
& & & \zz(\theta) \\ \vspace{8pt}
}
\end{equation*}
As identified above, the internal portion of the circuit corresponds to a ${ZZ}(\theta)$ interaction. This means that making $\mathcal{ZZ}(\theta)$ the compiler's native entangling operation instead of $\ms_\phi(\theta)$ provides opportunities to cancel some single-qubit gates. The standard decomposition of an arbitrary two-qubit gate due to the KAK decomposition is,
\begin{equation} \label{eq:kak}
\Qcircuit @C=.5em @R=.2em {
& \gate{A} & \multigate{1}{{XX}(t_x)} &  \multigate{1}{{YY}(t_y)} & \multigate{1}{{ZZ}(t_z)} & \gate{C} & \qw  \\
& \gate{B} & \ghost{{XX(t_x)}} &         \ghost{{YY(t_y)}} &        \ghost{{ZZ(t_z)}} & \gate{D} & \qw}
\end{equation}
where each two-qubit gate can be implemented with a single $\mathcal{MS}(\theta)$ or $\mathcal{ZZ}(\theta)$ application and surrounding single-qubit gates.
However, by compiling to $\mathcal{ZZ}$ as the native entangling operation, the co-propagating gates originally required for each $\mathcal{MS}$ will instead be merged with one another or with the outer single-qubit unitaries $\{A, B, C, D\}$ to decrease the overall depth of the circuit.

%


\subsubsection{{SWAP} mirroring}

{SWAP} Mirroring is built on the observation that there are instances when appending two {SWAP}s to an arbitrary two-qubit gate can be more efficient than performing the gate by itself. For example, transforming a double-{CX} circuit by appending two {SWAP}s (identity):
\begin{align*}
\Qcircuit @C=1em @R=.1em {
\lstick{q_0} &  \targ & \ctrl{1} & \rstick{q_0} \qw  &&& \push{\rule{0em}{2.8em}\rightarrow \rule{0em}{2.8em}} &&& \lstick{q_0} &  \targ & \ctrl{1} & \qswap &  \qswap & \rstick{q_0} \qw  \\
\lstick{q_1} & \ctrl{-1} & \targ & \rstick{q_1} \qw &&& &&&  \lstick{q_1} & \ctrl{-1} & \targ & \qswap \qwx & \qswap \qwx & \rstick{q_1} \qw}
\end{align*}    

\noindent The last {SWAP} can be performed ``virtually" by relabeling the qubits and the other {SWAP} can be decomposed into the {CX} basis:
\begin{align*}
\Qcircuit @C=.8em @R=.7em {
\lstick{q_0} & \targ & \ctrl{1} & \qw & \ctrl{1} & \targ & \ctrl{1} & \qw & \link{1}{-1} &  \rstick{q_1} \qw \\
\lstick{q_1} & \ctrl{-1} & \targ & \qw & \targ &  \ctrl{-1} & \targ & \qw & \link{-1}{-1} & \rstick{q_0} \qw} 
\end{align*}

\noindent The {CX} gates cancel twice to result in:
\begin{align*}
\Qcircuit @C=1em @R=.7em {
\lstick{q_0} & \ctrl{1} & \qw & \link{1}{-1} &  \rstick{q_1} \qw \\
\lstick{q_1} & \targ & \qw & \link{-1}{-1} & \rstick{q_0} \qw} 
\end{align*}

\noindent Thus, a two-operation circuit is optimized to a single operation.

{SWAP} mirroring is described more generally by leveraging Equation B9 in Appendix B of \cite{cross2019demonstration} which is described by the circuit:
\begin{align*}
\Qcircuit @C=.3em @R=.1em {
& \gate{A'} & \multigate{1}{{XX}(\frac{\pi}{4} - t_{x})} &   \multigate{1}{{YY}(\frac{\pi}{4} - t_{y})} &  \multigate{1}{{ZZ}(t_{z} - \frac{\pi}{4})} & \gate{C'} & \qw & \link{1}{-1} & \qw  \\
& \gate{B'} & \ghost{{XX}(\frac{\pi}{4} - t_{x})} &         \ghost{{YY}(\frac{\pi}{4} - t_{x})} &        \ghost{{ZZ}(t_{z} - \frac{\pi}{4})} & \gate{D'} & \qw & \link{-1}{-1} & \qw}
\end{align*}
Given an arbitrary two-qubit unitary, we use the KAK decomposition \cref{eq:kak} and compare it to decomposing via the above circuit. We determine which is the more efficient circuit first by whichever version yields a reduction in the number of M{\o}lmer-S{\o}rensen (MS) gates (both $\ms_{\phi}(\theta)$ and $\zz(\theta)$), and then by the least total MS angle subtended across all MS gates, $\theta_{MS}$, within the decomposition. Applying SWAP mirroring to Haar random 2-qubit unitaries yields 38 percent less $\theta_{MS}$ when successful and 17 percent average reduction in $\theta_{MS}$ in general circuits overall. This demonstration is relevant for applications like Supercheq fingerprinting \cite{gokhale2022supercheq} and quantum volume \cite{cross2019demonstration}.


\begin{figure}[h]
    \centering
    \includegraphics{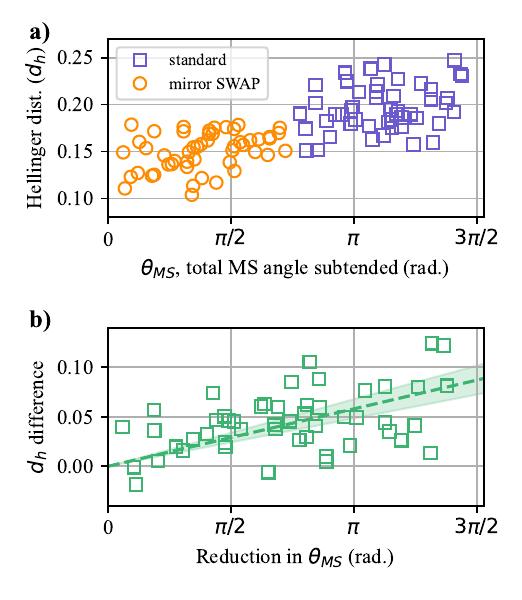}
    \caption{Experimental results for SWAP mirroring in the context of FSim($\Theta, \Phi$) gates on the QSCOUT trapped ion hardware. a) For fifty randomly chosen $\Theta$ and $\Phi$, the Hellinger distance, $d_h$, between the empirically observed and simulated ideal values for the output probability distribution is plotted versus $\theta_{MS}$, for both the standard compilation (blue) and the SWAP mirrored compilation (orange). b) For each FSim implementation, the difference between $d_h$ of the standard and SWAP mirrored compilations is plotted against the total reduction $\theta_{MS}$. Linear fit (dashed) with 2$\sigma$ error bars emphasizes trend of increased $d_h$ difference with increased $\theta_{MS}$ reduction.}
    \label{fig:mirrorSWAP_exp}
\end{figure}

On QSCOUT, we investigate SWAP mirroring as a means to reduce $\theta_{MS}$, within a circuit in order to improve its performance. For this demonstration, we use the  FSim($\Theta,\Phi$) gate and restrict ourselves to instances where $\pi/4<\Theta<3\pi/4$ and $\pi/2<\Phi<3\pi/2$. This range of parameter inputs of the FSim provides a region in which SWAP mirroring always provides a reduction in $\theta_{MS}$. We randomly choose 50 different variations of FSim parameters $\Theta$ and $\Phi$, and compile each FSim gate via the standard methodology or invoking SWAP mirroring. The circuit compilations consist of a series of $R_{\phi}(\theta)$ gates, virtual $Z(\theta)$ gates, and three $\zz(\theta)$ gates. In each case, we begin with an input state of $\ket{10}$, and measure the output state. The Hellinger distance, $d_h$ of the output state is computed in relation to the ideal output state of the circuit. In Fig.\ref{fig:mirrorSWAP_exp}a, we plot the $d_h$ of the output states of both the standard and SWAP mirrored compilation against the total MS angle subtended by the circuit, showing a clear improvement in the performance of the circuit under SWAP mirroring conditions. In Fig.\ref{fig:mirrorSWAP_exp}b, we look directly at the difference of $d_h$ per circuit and plot as a function of the reduction in $\theta_{MS}$, also showing a trend for greater differences at increased $\theta_{MS}$ reductions.  






\subsection{``Bring-Your-Own'' Gateset (BYOG)}

Any benefit from optimizing decompositions to a device's native gateset is ultimately limited by how well those native gates themselves can be implemented in the hardware. That is, the performance of the optimized circuit is critically dependent on the precise calibration of hardware operations to implement the gateset assumed by the compiler.
This accuracy is fundamentally limited by the precision and stability of the device and control hardware, as well as the budget available for calibration (which can be a source of considerable experimental overhead, as device and control system properties can be vary qubit-to-qubit and day-to-day). High-fidelity circuit execution therefore often relies on techniques such as echo sequences (discussed above), composite gates, and optimal control to improve native gate precision, at the cost of increased circuit complexity and runtime.

We therefore employ an alternative paradigm. Instead of compiling to an ideal gateset and then requiring it to be reproduced as well as possible in hardware, we first find a set of quantum operations which is easily implemented in the hardware, and then pass those to the compiler to use when decomposing logical gates. The task of gate calibration is then in part reduced to that of characterization. Spatial and temporal inhomogeneity is handled naturally: it can just be the case that the native gateset might include (for example) slightly different entangling operations on different pairs of qubits, all of which may be different than they were yesterday. The experimenter can focus on calibrating out unwanted factors (such as crosstalk \cite{ding2020systematic} to spectator qubits or leakage out of the logical subspace).

This compilation strategy comes at the cost of expressivity. It is generally the case that the uncalibrated gates will not generate arbitrary logical operations as efficiently as more conventional choices, which are typically chosen in part for this expressiveness.
This complexity can be reduced in part by employing approximate synthesis, in which logical operations are decomposed to sequences of native operations which are only required to approximate the desired unitary to within a provided tolerance (typically chosen to be just small enough that that approximation error does not contribute substantially to the overall error rate on the device).
We then find that the loss in expressivity is often overcome by the reduction in complexity of the native operations themselves: while the same logical operations require a shorter sequence of native operations given a more conventional gateset, the physical implementation of each of these native operations is typically more complex due to the extended or composite sequences which are necessary to precisely implement the chosen native operation. An example of this tradeoff will be shown in \cref{sec:qutrits}.

\subsubsection{Implementation}

We implement this strategy for two-qubit operations on the Advanced Quantum Testbed (AQT), a superconducting transmon quantum computer at Lawrence Berkeley National Laboratory. \Superstaq/'s compilation endpoint for AQT allows arbitrary unitary matrices to be assigned to pulse calibrations for the control hardware. \Superstaq/ will then assume the provided gate definitions to be the device's native gateset when compiling logical operations into physical pulse sequences.

We leverage the Berkeley Quantum Synthesis Toolkit (BQSKit) \cite{bqskit2021} for approximate synthesis. BQSKit provides powerful tools for fast synthesis given arbitrary basis operations. After merging adjacent two-qubit logical operations into a single unitary, we employ its ``QSearchSynthesis'' pass \cite{davis2020} to approximate the unitary to a provided precision using a sequence of two-qubit native operations and arbitrary single-qubit gates. The tolerance is chosen to balance approximation fidelity with that lost due to increasing sequence lengths.

Single-qubit operations are then decomposed analytically into AQT's native single-qubit $R_x(\pi/2)$ gates and virtual-$Z$ rotations. The decomposition chosen depends on the form of the provided unitary: if the virtual-$Z$ rotation on that qubit can be commuted through the subsequent two-qubit operation we use the ubiquitous $ZXZXZ$ decomposition; otherwise, we employ the PMW-4 sequence introduced in \cite{ding2022} to ensure that the virtual phase is returned to zero beforehand. In both cases, we employ optimizations like those described in \cite{hashim2022} to decrease the number of $R_x(\pi/2)$ pulse required for certain subsets of single-qubit unitaries (e.g. a Hadamard gate requires just one pulse in the former case and three in the latter).

\subsubsection{Extension to Qutrits} \label{sec:qutrits}

An immediate advantage to the BYOG model described above is that it is easily extended to higher-level systems by allowing for custom gate definitions of arbitrary dimension. Moving beyond the two-level (qubit) subspace is a significant way to expand the computational capability of a quantum device. By incorporating higher energy levels of the quantum systems, not only do we unlock an exponentially larger Hilbert space in which to perform quantum computations, but also higher effective connectivity than an equivalent system implemented on qubits. These factors have been shown to offer asymptotic speedups for important operations \cite{gokhale2019, shi2020resource, gokhale2020extending} given access to three-level (qutrit) operations.
The low anharmonicity of superconducting transmon qubits make them an ideal architecture for implementing qutrit gates, and indeed high-fidelity one- and two-qutrit quantum operations have already been demonstrated on AQT \cite{goss2022}.

The BQSKit approximate synthesis pass we employ supports qutrit gates out of the box. We decompose single-qutrit gates analytically, assuming a fixed single-qutrit gateset comprising $R_x(\pi/2)$ gates acting in the $\{\ket0,\ket1\}$ and $\{\ket1,\ket2\}$  subspace of SU(3), and arbitrary virtual-$Z$ rotations in either subspace. As in \cite{morvan2021}, arbitrary unitaries are first decomposed into a sequence three SU(2) operations in alternating two-level subspaces of SU(3). Each subspace operation is then deconstructed via the ubiquitous $ZXZXZ$ decomposition, using the available native operations acting in that subspace. The latter step allows for the same optimizations employed when decomposing single-qubit gates, in which each subspace operation may be implemented with zero or one $R_x(\pi/2)$ pulses if its rotation angle is sufficiently close to 0 ($\pm\pi/2$) (where "sufficiently close" can be configured to ensure a desired precision).

Finally, we note that this compilation strategy is compatible with Equivalent Circuit Averaging (ECA), another compilation technique implemented in \Superstaq/'s compilation endpoint for AQT to mitigate systematic errors (first introduced in \cite{hashim2022}). When using the BYOG compilation strategy, \Superstaq/'s ECA compilation endpoint exploits the stochasticity of the approximate synthesis pass to generate an ensemble of logically equivalent but physically distinct decompositions of each gate, which in turn are used to generate the set of equivalent circuits to average over.

\subsubsection{Demonstration}

\begin{figure}[htb]
    \centering
    \includegraphics[width=0.45\textwidth]{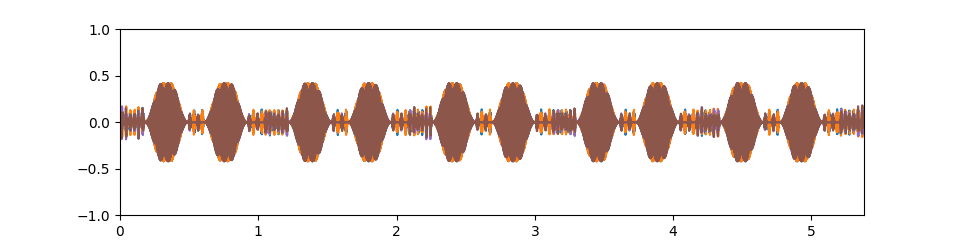}
    \includegraphics[width=0.45\textwidth]{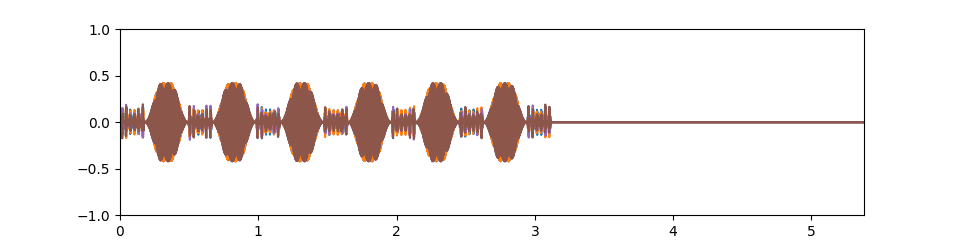}
    \caption{\label{fig:swap3-seqs} Pulse sequences implementing the same randomly-generated SU(9) operation on AQT hardware, decomposed using a conventional qutrit-CZ native operation (top) and the measured unitary of the cross-Kerr pulse alone (bottom). Though the latter requires a longer gate decomposition (with six applications of its native two-qutrit gate instead of five), the physical pulse sequence is shorter because it forgoes the echoing required to implement a conventional qutrit-CZ gate natively.}
\end{figure}

To demonstrate the efficiency of the BYOG compilation strategy, in \cref{fig:swap3-seqs} we present two different \Superstaq/-generated pulse sequences implementing the same randomly-generated SU(9) operation using two different native entangling gates. The first uses the precise qutrit-CZ operation described in \cite{goss2022}, which employs an echo sequence consisting of two cross-Kerr entangling pulses interleaved with $X$ gates in the $\{\ket0,\ket1\}$ subspace to annihilate unwanted entangling phases. In the second, we provide \Superstaq/ with just the pulse sequence and measured unitary of the cross-Kerr pulse itself. In both cases the tolerance used for approximate synthesis was set to $5\cdot10^{-4}$ (Hilbert-Schmidt norm, or equivalently a process fidelity of $10^{-3}$). We find that the qutrit-CZ is indeed more expressive, requiring just five qutrit-CZ operations to approximate the desired unitary compared to six two-qutrit operations when decomposed to the cross-Kerr pulse itself. However, because each qutrit-CZ gate itself comprises a sequence of two cross-Kerr interactions, this expressivity does not translate into a shorter physical implementation. As seen in \cref{fig:swap3-seqs}, the sequence length is reduced by about 42\% when we compile to the lower-level operation.

\begin{figure}[htb]
    \centering
    \includegraphics[width=0.4\textwidth, trim=0 30 0 60, clip]{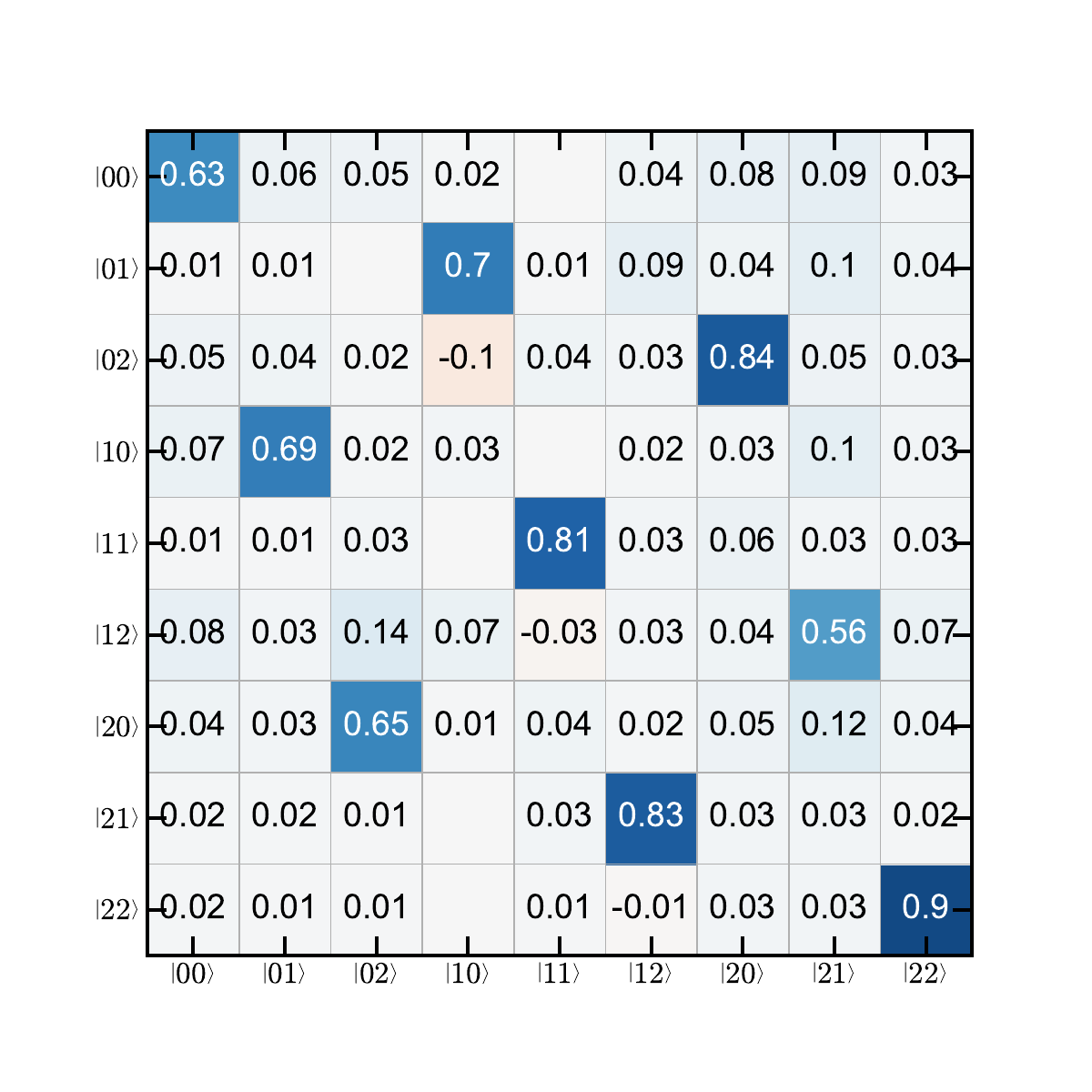}
    \caption{\label{fig:swap3-truth-table} Truth table measurement frequencies of a qutrit-SWAP operation (after readout correction). The overall truth-table fidelity was approximately 73.5\%.}
\end{figure}

Finally, we validate this compilation procedure experimentally by using it to demonstrate a two-qutrit SWAP operation. The SWAP circuit (including setup and measurement operations) is compiled using \Superstaq/'s public-facing API (interfaced via \texttt{cirq-superstaq}), to which we provide the pulse definitions and measured unitary of the aforementioned cross-Kerr interaction and pulse definitions for the relevant single-qubit native operations. The pulse sequences returned by \Superstaq/'s compilation endpoint for AQT are then implemented directly on the AQT hardware.
Readout correction is applied by measuring and inverting the confusion matrices (cf. \cite{bravyi2021}).
The resulting measurement frequencies for all computational-basis input and output states are shown in \cref{fig:swap3-truth-table}, from which we find an overall truth-table fidelity of about 73.5\%. Note we do not employ ECA for this demonstration.

\section{Dynamical Decoupling} \label{sec:dynamical_decoupling}

\subsection{Introduction}
\label{sec:dd_introduction}

Noisy hardware is an important consideration when building a quantum compiler.
Through cross layer optimization, \Superstaq/ attempts to create performance gains at every layer of the stack.
To this end, one of the most important and promising low-level techniques for quantum error suppression is known as dynamical decoupling (DD) \cite{viola1999dynamical, ezzell2022dynamical}.
At a high level, DD works by injecting \emph{additional} operations that are engineered to suppress the buildup of coherent errors in a circuit.
In a noiseless setting, the operations injected by DD would cancel each other out and resolve to the identity.
For this reason, a compiler designed merely to simplify and shorten circuits would eliminate DD operations, resulting in worse overall performance.
\Superstaq/ balances the objectives of shortening circuits with the benefits of DD, and has built a DD suite to target a variety of hardware backends.

The idea behind DD is that coupling between qubits and their environment can lead to undesired qubit evolution, such as stray rotation by an unknown angle on the Bloch sphere.
These processes destroy the quantum information stored in a qubit.
DD addresses these errors by applying regular pulse sequences that can be thought of as changing the ``frame'' in which the qubit stores information, e.g. by regularly swapping the north and south poles of the Bloch sphere.
In the co-rotating frame of the qubit, the unknown environmental couplings then average out to zero, thereby eliminating the buildup of coherent error.

The simplest DD sequence is known as CPMG \cite{galitski2013spin, bromborsky2014introduction}, which simply applies periodic, evenly-spaced $X$ gates, known as $\pi$-pulses, to a qubit.
However, this sequence does not protect against stray $R_x$ rotations, which commute with (and are therefore unaffected by) the CPMG pulses.
This limitation can be fixed simply by alternating the axis of rotation $X$ and $Y$, resulting in the so-called XY4 sequence (because four rotations are necessary for a single ``period'' that brings the qubit back to its original state).
XY4 is thus the simplest \emph{universal} DD sequence, which is to say that it can mitigate stray qubit rotations about any axis.
Making further modifications to XY4, for example by using additional axes of rotation, recursively nesting DD sequences into the gaps between DD pulses, or using unevenly spaced DD pulses, yields a large family of DD sequences that exhibit different advantages in the presence of different environmental or control errors \cite{ezzell2022dynamical}.
In all cases, DD works best when a compiler has control over the timing of the pulses that it injects into a circuit.

\begin{figure}
    \centering
    \includegraphics[width=0.4\textwidth]{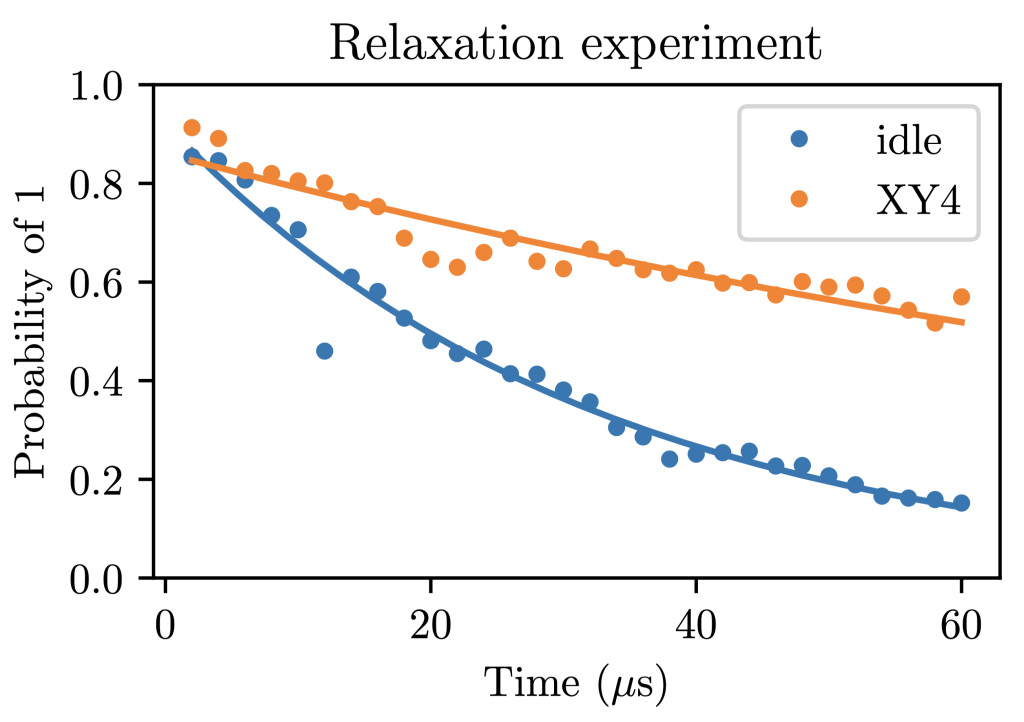}
    \caption{
    After preparing a qubit in the state $\ket{1}$, interactions with the environment cause the qubit to lose energy and decay to $\ket{0}$ as it idles.
    However, inserting an XY4 DD sequence mitigates this process and extends the qubit lifetime.
    Lines show a fit to an exponential decay curve.
    Figure reproduced with permission from Ref.~\cite{aws_rigetti}.
    }
    \label{fig:T1Rigetti}
\end{figure}

As a simple demonstration of the benefits of DD, \cref{fig:T1Rigetti} shows the results of an experiment to measure the relaxation ($T_1$) time of an idling qubit on Rigetti's Aspen M-3 quantum processor. We first initialize a qubit in the $\ket{1}$ state, and show the probability that the qubit is still observed in $\ket{1}$ upon measurement at a later time.
We either let the qubit idle between state preparation and measurement, or we insert two evenly-spaced XY4 repetitions (8 pulses total) during the idle time.
Altogether, inserting this DD sequence increased the lifetime of a qubit by a factor of 4. 

\subsection{Execution}
An effective DD framework must have the ability to employ both (a) the diverse set of DD sequences referenced above and (b) various levels of concatenation during a period of qubit idling.
Further, DD must have context of both program and machine properties during runtime for error mitigation to be most effective~\cite{smith2022}. For example, many factors influence the noise a qubit encounters as it idles during program execution, causing some implementations of DD to be more effective than others. Exemplar factors include: duration of qubit idle windows (how many DD sequences can we fit and is there optimal spacing between the DD operations?); circuit information (what state is the idling qubit holding?); qubit coherence properties (will one type of DD correction be better than another for a qubit?); operator error rates (must balance tradeoffs between decoherence resilience and extra gate-induced error that each DD operation injects); parallel gate executions (does crosstalk affect an idling qubit?) native gate-set (what DD sequence is easiest to implement with built-in operations?); and architectural constraints of classical control hardware (how precise is the classical control infrastructure?). These factors simultaneously influence the optimal type of DD, in terms of operator sequence chosen and number of iterations within an idle window, that provides a quantum program with the greatest noise resilience during execution.

Quantum hardware with low level access typically provides a mapping of instructions to sequences of low-level control signals (pulses) that carry out the target instruction. The \Superstaq/ DD optimizer uses this mapping to schedule circuits. In particular, when a circuit is scheduled, every circuit operation is annotated with its execution start time and duration. This timing data allows us to compute important circuit information such as a circuit's critical paths and periods of qubit idling. For each qubit in an operation, we use the timing data to determine the start and end times of the qubit's idling periods. These idling durations are the points in the circuit that we target with appropriately spaced DD sequences, such as CPMG (XX), XY4, and XY8\cite{souzaAl}, using individual and concatenated pulse sequences. Our implementation is written in a generalizable fashion that can flexibly accommodate other DD sequences as well.

We also take quantum computing hardware timing constraints into consideration when scheduling a circuit. If a circuit does not satisfy the system's timing criteria in terms of thresholds for pulse duration, alignment, and granularity, the scheduled circuit will not be executable because the circuit instructions cannot be realized with the low-level control hardware. An example of low-level control hardware would be the arbitrary waveform generators (AWGs) that drive superconducting transmon qubits (i.e. IBM's qubit technology). IBM's quantum hardware currently requires that pulses begin at a time that is a 16-fold multiple of a specified (by the device's timing constraints) alignment value expressed in device-dependent timescale increment, \textit{dt}\cite{backendinfoqiskit}. To respect this constraint, when compiling circuits targeting IBM's hardware we align each operation $i$ to begin at $16k_i$ \textit{dts} for a whole number $k_i$.

\subsection{Results}

\begin{figure}[h]
    \centering
    \includegraphics[width=0.4\textwidth,trim=25 30 25 30]{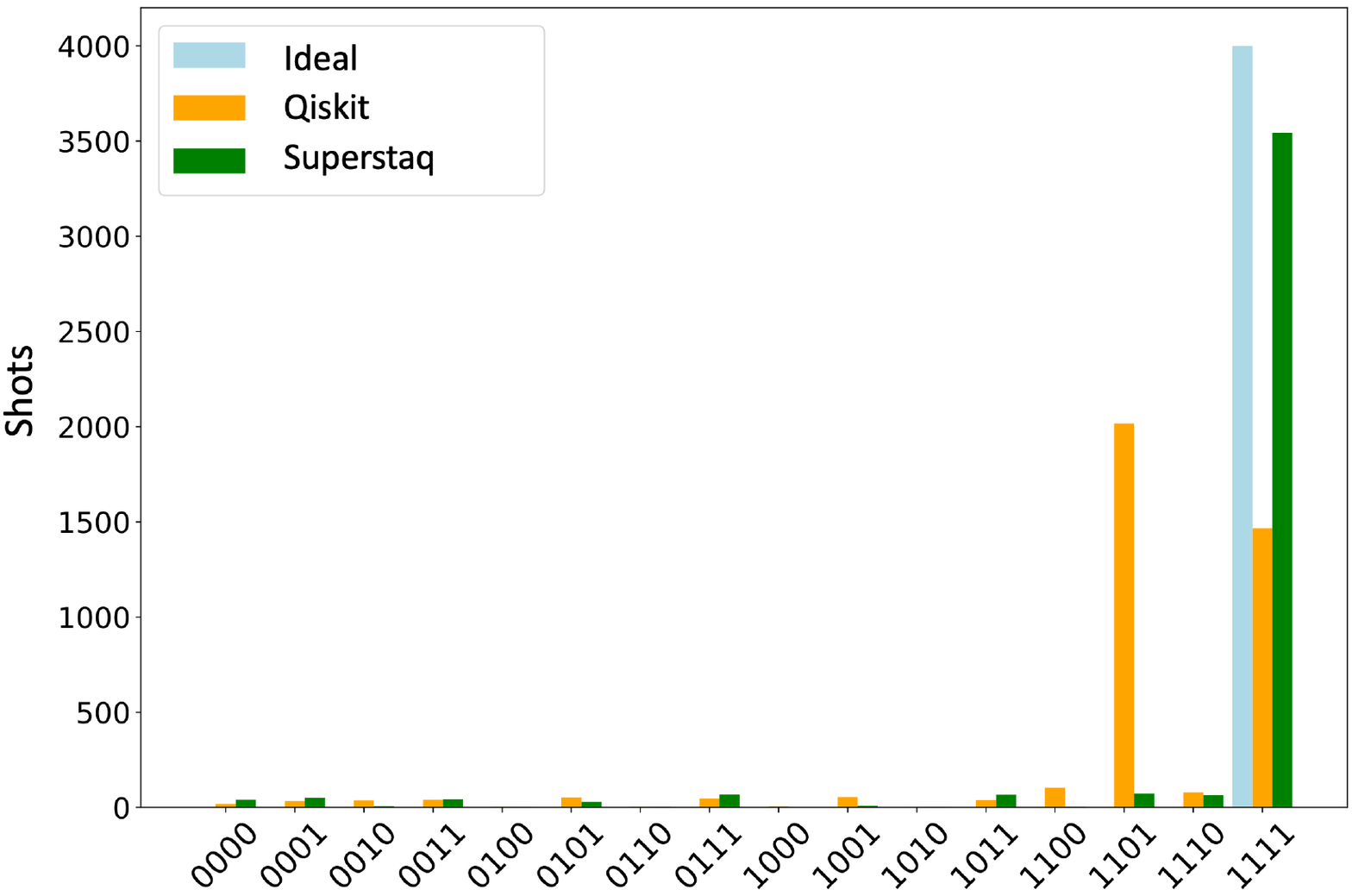}
    \caption{DD results on IBM Hanoi quantum computer for 4-qubit Bernstein-Vazirani  circuit with (`1111' key).}
    \label{fig:DD_IBM_case_study}
\end{figure}

Fig.~\ref{fig:DD_IBM_case_study} shows how DD makes algorithm execution more noise-robust. In this case study, we ran 4000 shots of a four-qubit implementation of the Bernstein-Vazirani (BV) benchmark (secret string = `1111') on the 27-qubit IBM Hanoi device. Fig.~\ref{fig:DD_IBM_case_study} compares a level-3 Qiskit optimization and \Superstaq/ DD optimization to the ideal single-peak distribution. We see that invoking \Superstaq/ DD improves the fidelity to the ideal distribution from $37\%$ to $89\%$. Not only did \Superstaq/ DD boost the likelihood of observing the correct all-1's outcome for the BV application, \Superstaq/ DD made a previously unobtainable distribution feasible on the same hardware.

\begin{figure}[h]
    \centering
    \includegraphics[width=0.42\textwidth, trim=15 5 15 15, clip]{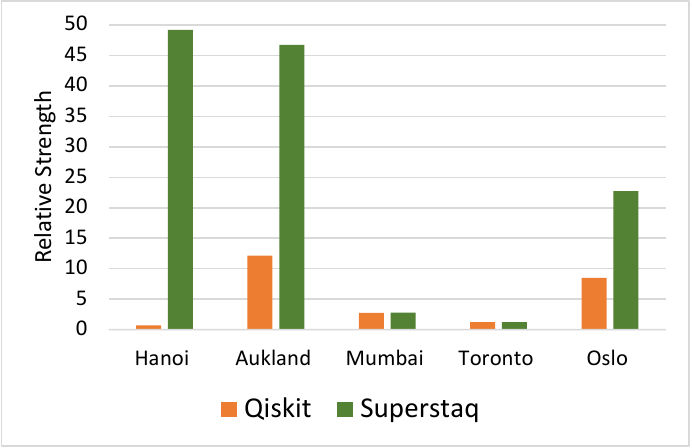}
    \caption{DD results across five IBM machines for a four-qubit Bernstein-Vazirani program. \Superstaq/ increases relative strength as high as 68x relative to Qiskit with \texttt{optimization\_level=3}.}
    \label{fig:DD_IBM_rs}
\end{figure}

The \Superstaq/ DD optimizer was designed to encompass the ``write once target all'' objective of \Superstaq/. It complements other circuit optimization passes, such as gate-count minimization and noise aware mapping, to produce DD scheduling that is customized for an algorithm and QC paring.
In Fig.~\ref{fig:DD_IBM_rs}, \Superstaq/ DD is compared to Qiskit's highest optimization level for five different machines using the four-qubit BV benchmark. To better understand how a circuit optimization mitigates noise and sharpens a distribution, the figure of merit known as relative strength \cite{tannu2020software} is often used. Relative strength for a distribution is defined as the ratio of total correct observations to most frequent incorrect observations. \Superstaq/ DD significantly improves program outcomes, with results as high as 68x on the Hanoi machine relative to the Qiskit baseline. Although \Superstaq/ DD had about the same relative strength as Qiskit on Mumbai and Toronto, we still observed improvements in the probability of success through \Superstaq/. 


\section{Star-to-Line Routing} \label{sec:advanced_mapping}

Often, the connectivity graph of target quantum circuits do not match the connectivity graph of target quantum hardware. Therefore, mapping and routing is required to execute the quantum circuit. Mapping refers to the initial assignment of physical qubits on the device to qubits in the quantum circuit. Routing refers to the process of inserting SWAPs that enable interaction between non-local physical qubits.

\begin{figure}[h]
    \centering
    \includegraphics[width=0.5\textwidth]{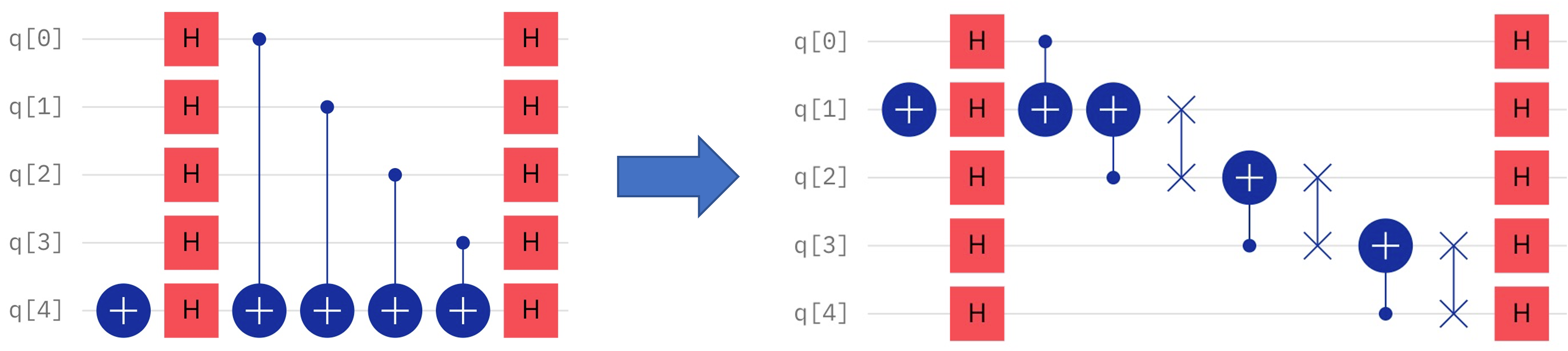}
    \caption{Star-to-line mapping used for five-qubit BV algorithm.}
    \label{fig:star2line}
\end{figure}

Many target quantum algorithms, such as Bernstein-Vazirani (BV), the Quantum Fourier Transform (QFT), and implementations of the variational quantum eigensolver (VQE)
, are dominated by instances of \emph{star} connectivity in their program structure where a single qubit interacts with every (or nearly every) other qubit. Unfortunately, many quantum hardware platforms---especially superconducting---exhibit sparse linear connectivity between physical qubits. This motivated the development of the \emph{star-to-line} routing pass, seen in Fig.~\ref{fig:star2line} for the five-qubit BV algorithm.

As depicted, the algorithm has a star connectivity, whereby q[4] interacts with every other qubit. This star connectivity can be effectively converted to a linear connectivity circuit with modest overhead. In particular, the CX between the second-to-edge qubit (q[1]) and edge qubit (q[0]) can be implemented without any routing. Then a CX and SWAP are applied between the second-to-edge qubit (q[1]) and its other neighboring qubit (q[2]). This process continues, applying a CX and SWAP between physical qubits q[i] and q[i+1], until we reach the end of the line. Moreover (not shown for brevity), the remaining sequences of CX-SWAP can be performed with just two CX gates, via standard gate cancellation identities \cite{tomesh2021coreset, tomesh2022supermarq}.

\begin{figure}[h]
    \centering
    \includegraphics[width=0.4\textwidth]{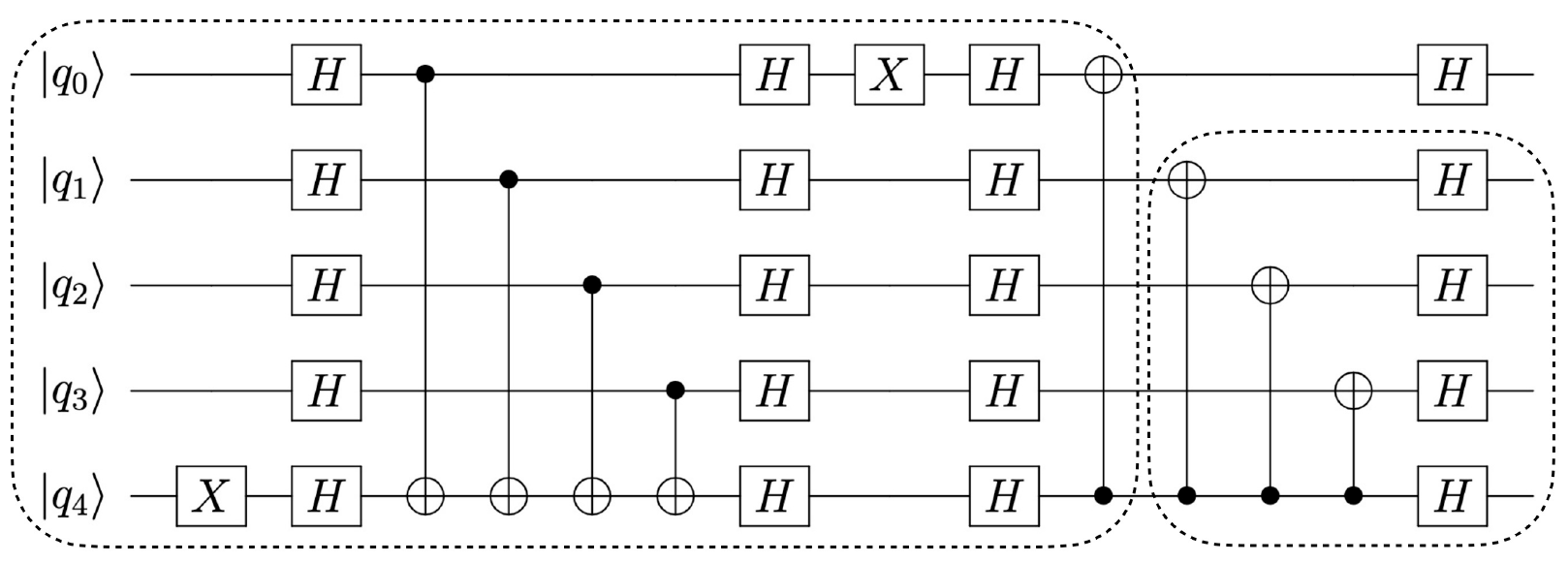}
    \caption{Example of a circuit which contains star-connected sub-circuits.}
    \label{fig:star_finder}
\end{figure}

\Superstaq/'s star-to-line router parses through a given quantum circuit to find sub-circuits which have star connectivity. \Cref{fig:star_finder} provides an example of such a circuit. The dashed lines indicate sub-circuits which have star connectivity.

\begin{figure}[h]
    \centering
    \includegraphics[width=0.42\textwidth]{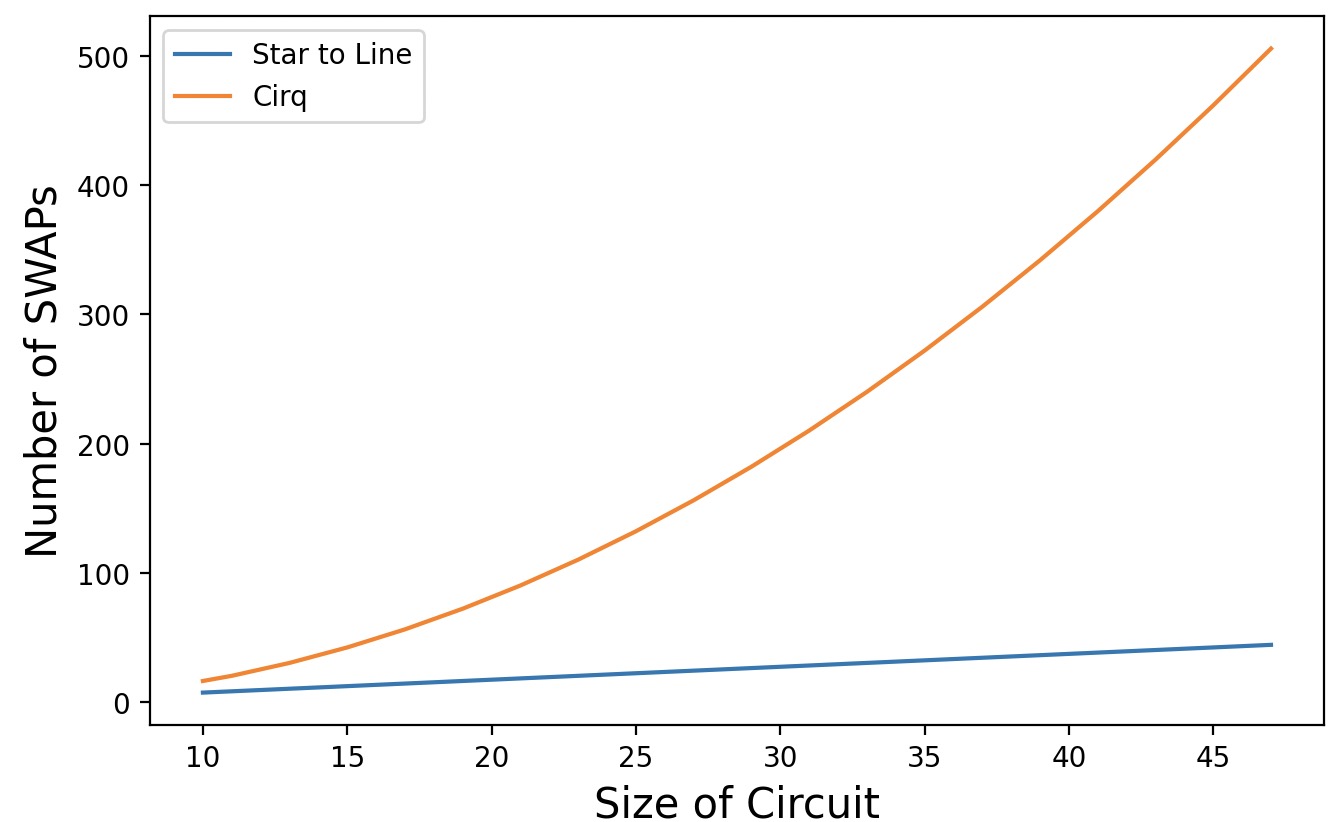}
    \caption{Number of SWAPs vs length of circuit for \Superstaq/'s star-to-line router and \texttt{Cirq}'s router.}
    \label{fig:cx_graph}
\end{figure}

\noindent We find the star-to-line router outperforms other state-of-the-art routers. Fig.~\ref{fig:cx_graph} plots the number of SWAP gates required to map a Bernstein-Vazirani circuit to a linear topology. The SWAP count increase quadratically with the \texttt{Cirq} router, but just linearly with \Superstaq/'s router, as expected.

\section{Conclusion}

The premise of \Superstaq/ is that we can extract greater quantum program performance with deep cross-layer optimization tailored to the underlying hardware. In the process of building \Superstaq/, we advanced techniques and insights related to parametric (fractional) gates, dynamical decoupling, swap mirroring, bring-your-own gateset, Phased MicroWave decompositions, approximate synthesis, and qutrits---all of which are presented in this paper. We also find it profitable to design compilation with typical application workloads in mind, which motivates the star-to-line mapping technique. Overall, \Superstaq/ is able to achieve significant improvements in quantum program performance, exemplified for instance by our $>$ 10x performance  improvements for benchmark applications. We hope that the open-beta availability of \Superstaq/, through its open-source \texttt{qiskit-superstaq} and \texttt{cirq-superstaq} clients, will enable practitioners and researchers to continue to advance the frontier of what quantum computers can accomplish.


\section*{Acknowledgments}
We thank Woo Chang Chung, Dan Cole, David Mason, Tom Noel, and Alex Radnaev of Infleqtion for their support towards executing quantum circuits on Hilbert.

\bibliographystyle{unsrt}
\bibliography{refs}

\end{document}